\documentclass{pnastwo}
\usepackage[dvips]{graphicx}
\usepackage{pnastwoF}
\usepackage{epsfig}
\usepackage{bm}
\def\vec#1{\bm{#1}}

\begin{document}

\conflictofinterest{Conflict of interest footnote placeholder}

\track{This paper was submitted directly to the PNAS office.}

\footcomment{Abbreviations: FvK, F\"oppl-von K\'arm\'an  number }
\title{Mechanical~limits~of~viral~capsids}
\author{Mathias Buenemann\affil{1}{Fachbereich Physik,
    Philipps-Universit\"at Marburg, D-35032 Marburg}, Peter
  Lenz\affil{1}{}\affil{2}{Center for Theoretical Biological Physics, University of
  California at San Diego, La Jolla, California 92093-0374, USA} }


\contributor{Submitted to Proceedings of the National Academy of Sciences
of the United States of America}

\maketitle

\begin{article}

\begin{abstract}
  We study the elastic properties and mechanical stability of viral
  capsids under external force-loading with computer simulations.  Our
  approach allows the implementation of specific geometries
  corresponding to specific phages such as $\phi29$ and CCMV. We
  demonstrate how in a combined numerical and experimental approach
  the elastic parameters can be determined with high precision.  The
  experimentally observed bimodality of elastic spring constants is
  shown to be of geometrical origin, namely the presence of
  pentavalent units in the viral shell.  A criterion for capsid
  breakage is defined, which explains well the experimentally observed
  rupture.  From our numerics we find for the dependence of the
  rupture force on the F\"oppl-von K\'arm\'an (FvK) number $\gamma$ a
  crossover from $\gamma^{2/3}$ to $\gamma^{1/2}$. For filled capsids
  high internal pressures lead to a stronger destabilization of
  viruses with a buckled ground state than unbuckled ones. Finally, we
  show how our numerically calculated energy maps can be used to
  extract information about the strength of protein-protein
  interactions from rupture experiments.
\end{abstract}

\dropcap{B}acteriophage capsids have astonishing elastic properties.
They withstand extreme internal pressures exerted by their densely
packed DNA. DNA packaging experiments on phage $\phi29$ \cite{smith01}
and theoretical arguments
\cite{riemer78,kindt01,tzlil03,purohit03,purohit05} imply that phage
capsids store their DNA under $\sim$50atm. This pressure is necessary
to inject the DNA into the prokaryotic host cell \cite{atlas}. It has
been shown experimentally that the ejection pressure from
$\lambda$-phages is $\sim$20atm \cite{evilevitch03}.  The remarkable
robustness under high pressures motivated recent nanoindentation
studies \cite{wuite04}, which showed that capsids also resist external
point forces up to $\sim$1nN.

In contrast, viruses that infect eukaryotic cells, e.g. CCMV and
polyoma, penetrate their host cell.  Their DNA is released by
subsequent disassembly of the shell. Correspondingly, these viruses do
not store their DNA under high pressure.  Nevertheless, their
resistance to external forces is remarkably strong \cite{wuite06}.

Viral capsids are composed of a small number of different proteins,
which cluster to morphological units (``capsomers'').  These units are
put together in a highly regular fashion described by the
``quasi-equivalence'' principle by Caspar and Klug
\cite{casparklug62}.  Due to the presence of capsomers the surface of
capsids has a discrete structure. Mathematically, these capsomers
correspond to the vertices of a regular triangulation of a sphere. For
such a triangulation the number of vertices $V$, connecting edges $E$,
and associated faces $F$ have to fulfill Euler's theorem $F-E+V=2$.
This theorem implies that capsids of icosahedral symmetry have 12
pentavalent morphological units (``pentamers'') embedded in an
environment of hexavalent units (``hexamers'').

Viral capsids can undergo elastic and bending deformations. Generally,
the elastic properties of such thin shells (of typical length scale
$R$) depend only on a single dimensionless parameter, the FvK number
$\gamma \equiv R^2 \kappa_e/\kappa_b$ set by the ratio between elastic
modulus $\kappa_e$ and bending rigidity $\kappa_b$.  It has been shown
in Ref.  \cite{lidmar03}, that at a critical $\gamma_b=154$ spherical
shells undergo a buckling transition, in which the shell acquires a
more faceted shape. This transition is caused by large strains
associated with the pentamers which are reduced upon buckling into a
conical shape.

Typical $\gamma$ values of viral shells lie in the range from below
the buckling threshold (e.g. alfalfa mosaic virus with $\gamma\simeq
60$ \cite{kumar97}) up to several thousand (e.g. phage $T4$ with
$\gamma\simeq 3300$ \cite{olson01}).  An extreme example is the giant
mimivirus (diameter of $\simeq600$nm) with a Fvk number $\gamma\simeq
20000$ \cite{xiao05}. Even higher $\gamma$-values are possible for
artificial capsules such as vesicles with crystallized lipid membrane
\cite{dubois01} ($\gamma\sim10^5$) or polyelectrolyte capsules
\cite{elsner06} ($\gamma\sim10^6$).

The elastic properties of capsids can be probed in scanning-force
microscopy (SFM) experiments \cite{wuite04,wuite06}.  Depending on the
strength of loading two regimes are explored. Small forces lead to a
completely reversible deformation of capsids. Here, the linear and
nonlinear regime of thin spherical shell elasticity can be studied.
In this regime the deformations explore the {\sl global} elastic
properties.  Larger forces ($>$1nN) cause irreversible changes in the
shell structure commonly attributed to bond rupture.  Rupturing
studies therefore give information about the molecular interactions
between capsomers thus elucidating {\sl local} features of shell
mechanics.

Viral shells have also been the subject of numerical investigations.
The dependence of virus shape on the FvK number was analyzed in Ref.
\cite{lidmar03}.  Only recently, the elasticity of capsids has come
into the focus of numerical studies \cite{gompper06} which are based
on a discretization scheme introduced for crystalline membranes in
Ref.  \cite{seung88}.  In other approaches, the shells are directly
built up of proteins \cite{reddy98} or capsomers \cite{zandi04} with
specified (spatially varying) interactions.  On this basis the
stability of capsids against internal pressure has been studied
\cite{zandi05}.

Here, we generalize a numerical approach developed for the
investigation of vesicles \cite{wintz97} to viral capsids.  The big
advantage of our method is its applicability to arbitrary geometries.
In particular, our discretization of the bending energy does not
depend on the underlying triangulation of the viral surface.
Therefore, our method produces highly stable and reliable results even
under high local strain which allows us to investigate the rupture of
mechanical shells. In our simulations, we can systematically vary the
elastic moduli and geometry of the capsids and probe their mechanical
response to external disturbances. We can even implement specific
geometries, corresponding to specific phages and viruses, and
determine (by a direct comparison with experimental indentation
experiments) with high precision their elastic parameters, such as
linear and nonlinear spring constants, and the FvK number.  In our
simulations, we are also able to access features which are not
observable in experiments.  For example, by measuring the local strain
we are able to determine numerically the spatial distribution of
rupture probabilities across the capsid surface.  We will show that
experimental deviations from this distribution may be used to draw
conclusions about the spatial variation of protein binding strength.

This paper is organized as follows: after a short summary of the
numerical methods, we first study the global response of capsids to
externally applied point forces.  The numerical results for (empty)
$\phi$29 and CCMV are compared with experimental data. Next, we
analyze the local, irreversible response to indentation forces and we
discuss the dependence of local rupture-probability on the geometry of
the capsid.  Finally, our analysis is extended to filled capsids whose
elastic parameters and mechanical stability is studied in the last 2
sections.

\section{Methods}

\begin{figure}
  \begin{center}
    {\epsfig{figure=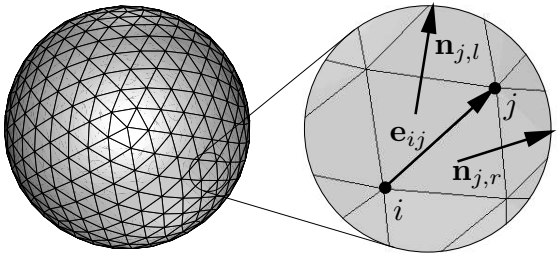,width=7.5cm}} 
  \end{center}
  \caption{\label{fig:model} The viral structure is represented by a
    triangulated sphere.  The pentavalent vertices represent the
    centers of the pentameric units.  Vertices are connected by
    elastic bonds with spring constant $\tilde{\kappa}_e$. The local
    mean curvature is calculated from the normals of neighboring
    facets and their shared edge.}
\end{figure}
We have performed numerical minimization simulations of triangulated
surfaces representing the surface of capsids with elastic (stretching)
and bending energy.  In our (small mesh-size) triangulation every
capsomer is represented by several vertices.  Therefore, these units
also have some flexibility and elasticity.  Such a discretization is
suitable to determine the shape of viral shells \cite{lidmar03} and
yields results which do not depend on the number of vertices (in
contrast to coarse triangulations in which each capsomer is
represented by a single vertex \cite{gompper06}).  The triangulation
represents the underlying icosahedral (quasi-)symmetry of the capsid.
In particular, the pentavalent vertices correspond to the centers of
the pentamers.

In our model, the vertices are connected by harmonic
springs. Any deviation from the preferred inter-vertex distance
$l$ gives rise to an elastic energy \cite{seung88},
\begin{equation}\label{eq:es_disc}
  E_e =\frac{\tilde{\kappa}_{e}}{2} \sum_{\langle i,j \rangle}
  (l-|\vec{r}_i-\vec{r}_j|)^2, 
\end{equation}
where vertex $i$ is at position $\vec{r}_i$ and the sum extends over
nearest neighbor pairs only.  $\tilde{\kappa}_e$ is related to the 2D
Young modulus $\kappa_e$ via $\tilde{\kappa}_e=\sqrt{3}\kappa_e/2$.

The bending energy is given by
\begin{equation}
  \label{eq:eb_disc2}
  E_b=\kappa_b\sum_i \frac{H_i^2}{A_i}, 
\end{equation}
where the sum extends over all vertices $i$. Here, $A_i$ is the area
assigned to the $i$-th vertex given by 1/3 of the area of all adjacent
triangles.  $H_i$ is the mean curvature of vertex $i$ (with
coordination number $z_i$) given by \cite{wintz97}
\begin{equation}
  H_i = \frac{1}{2}\sum_{j=1}^{z_i} \frac{\vec{e}_{ij}\cdot 
    (\vec{n}_{j,l}\times\vec{n}_{j,r})}{|\vec{n}_{j,r}||\vec{n}_{j,l}|+\vec{n}_{j,r}\cdot\vec{n}_{j,l}}, 
\end{equation}
with $\vec{e}_{ij} \equiv \vec{r}_i-\vec{r}_j$, and $\vec{n}_{j,l}$
and $\vec{n}_{j,r}$ are the non-normalized surface vectors of the
facets left and right of the edge connecting vertex $i$ and $j$, see
Fig. \ref{fig:model}. 

Starting from a triangulated sphere the shape of minimal energy is
found by minimizing the total energy $E=E_e+E_b$ using a conjugate
gradient algorithm \cite{press92}.  In order to simulate the
SFM-experiments mentioned in the introduction vertices $i$ at the
point of loading were moved away from their equilibrium position
$\vec{r}_i$. We thus work in an ensemble of prescribed indentation
$\vec{X}$ rather than in an ensemble of applied force. All vertices
are constrained to lie above a (virtual) plane at $z=0$. The shape of
the triangulated surface under these constraints is again found by
minimizing the total energy.

The shapes of bacteriophage $\phi$29 and the CCMV plant virus have
been quantitatively characterized by Cryo-EM and X-ray studies
\cite{speir95,tao98}.  Phage $\phi$29 has an average equatorial radius
$R\simeq 21$nm and a shell thickness of $h\simeq 1.6$nm. Its FvK
number is $\gamma \approx 11(R/h)^2 \simeq 2000$ implying that the
shape of $\phi$29 is noticeably buckled.  For CCMV the corresponding
values are $R\simeq 12$nm and $h\simeq 2.5$nm. Thus, $\gamma\simeq
300$ and CCMV is slightly buckled.

\section{Results and Discussion}

{\bf Spring Constants.} As mentioned in the introduction loading with
small forces probes the global elastic behavior of viral shells.
Here, we analyze the influence of internal structure on the elastic
response of capsids by comparing the numerical simulations with
analytical results for small deformations. Furthermore, we extract the
values of the elastic moduli of $\phi29$ and CCMV by a direct
comparison of numerical and experimental force-distance relations.

In a first step we have numerically calculated force-distance curves
for viruses indented by a distance $X \equiv \xi R$ for $0 \leq \xi
\leq 0.4$ (varying with step-size $\Delta \xi=1/250$). By minimizing
the total energy the (dimensionless) force $F$ is obtained by
$F=-\Delta E/(\kappa_b \Delta\xi)$, where $\Delta E$ is the change in
total energy caused by the increase $\Delta\xi$ in $X$.  Both hexamers
and pentamers were displaced. The indented hexamers discussed here all
lie in the center of the facet spanned by their three neighboring
pentamers.

\begin{figure}[bt]
  \begin{center}
    {\epsfig{figure=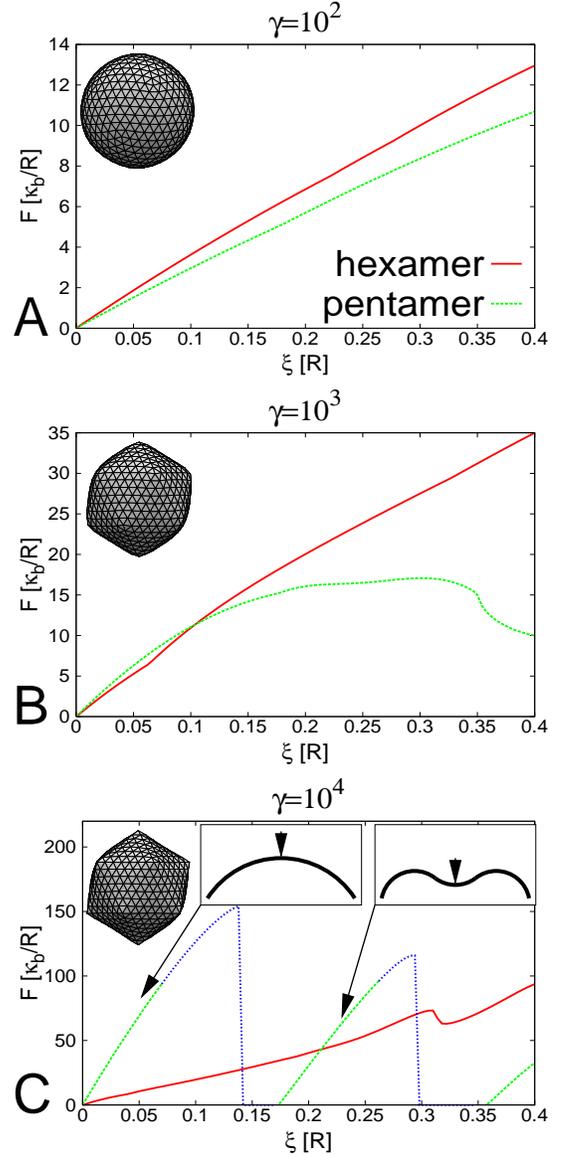,width=7.15cm}} 
  \end{center}
  \caption{\label{fig:1} Force-distance relations of locally deformed
    viral shells. For small $\gamma$ the force-distance relation is
    linear $F \sim \xi$ for the full range of applied forces (A).
    Above the buckling threshold the structural inhomogeneity of the
    capsid material is reflected by the increasing differences in the
    elastic response of pentamers and hexamers (B) and (C).  Above
    $\gamma\approx 1000$ the inversion transition of pentamers (see
    Fig. \ref{fig:4}B) causes softening, which leads to a decrease in
    $F(\xi)$. The influence on hexamers is much weaker.  At higher
    $\gamma$ pentamers may even snap into a new stable inverted
    configuration, which leads to discontinuities of $F(\xi)$.  The
    dotted blue line in (C) represents metastable conformations which
    have a higher energy than stronger deformed ones. Hexamers do not
    show a discontinuous inversion transition.}
\end{figure}
\begin{figure}
  \begin{center}
    {\epsfig{figure=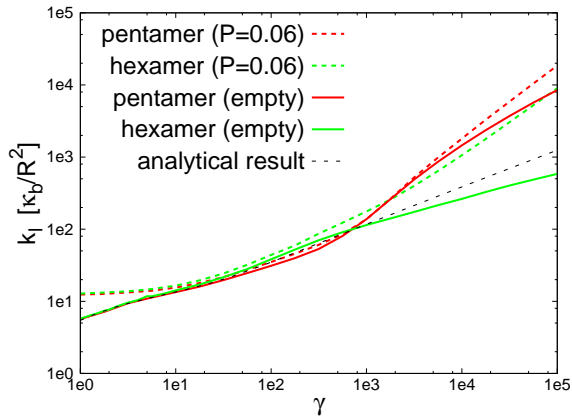,width=7.5cm}} 
  \end{center}
  \caption{\label{fig:3} Linear spring constants as a function of
    $\gamma$ for empty (solid lines) and filled (dashed lines)
    icosahedral capsids. The black dashed line represents the
    analytical solution (see suppl.  mater.). For $\gamma<\gamma_b$
    the spring constants of the empty capsid follow closely the
    theoretical prediction.  Above $\gamma_b$, the spring constant of
    hexamers $\sim \gamma^{1/3}$.  Pentamers become stiffer and
    (roughly) $\sim\gamma^1$.  Generally, filled capsids have higher
    spring constants. Data for the filled capsids is for
    $p=0.06\kappa_e/R$ (internal pressure of phage $\phi29$).}
\end{figure}

As can bee seen from Fig. \ref{fig:1} the elastic response of the
capsid to the deformation strongly depends on the ratio of bending and
elastic energy characterized by the FvK number $\gamma$.  For small
$\gamma$ capsids have a nearly spherical shape \cite{lidmar03} and
behave like homogeneous continuous shells.  For $10<\gamma<500$ the
(dimensionless) linear spring constants $K_l$ of both hexameric and
pentameric regions follow a square root law for sufficiently small
$\xi<\xi_c(\gamma)$
\begin{equation}
  \label{eq:4}
  F = K_l\xi \equiv 4\sqrt{\gamma}\xi\ , 
\end{equation}
see Fig. \ref{fig:3}.  The transition to the nonlinear regime (at
$\xi=\xi_c$) takes place at smaller $\xi_c$ the larger $\gamma$, see
Fig.  \ref{fig:1}A and B.  For small FvK number $\gamma \sim (R/h)^2$
corrections to thin shell theory of order $h/R$ become relevant.

For $\gamma>500$ the capsid has a strongly faceted shape since the 12
disclinations buckle.  Due to this structural inhomogeneity the
elastic response of pentamers and hexamers is different, i.e.
pentamers generally become stiffer than hexamers with increasing
$\gamma$. Furthermore, the abrupt shape transformation shown in Fig.
\ref{fig:1}C only occurs for loading on pentamers.

As Fig. \ref{fig:3} shows, the difference between the spring constants
for pentamers and hexamers increases with increasing $\gamma$. At
$\gamma \approx 2000$ the spring constants are already well separated
explaining the experimentally measured bimodality of spring constants
of $\phi 29$ \cite{wuite04}. From our numerical data we find $K_l \sim
\gamma^{1/3}$ for hexamers and $K_l \sim \gamma^1$ for pentamers.
When loading on pentamers is parallel to ridges then
$K_l\sim\gamma^{5/6}$ in agreement with the findings of Refs.
\cite{lobkovsky97,didonna01}.

\begin{figure}
  \begin{center}
    {\epsfig{figure=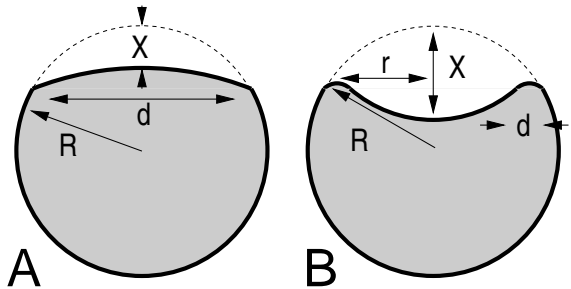,width=7.5cm}} 
  \end{center}
  \caption{\label{fig:4} Global response of elastic spherical capsids
    to local deformations. (A) For small indentations only the top of
    the shell is flattened giving rise to a linear force-distance
    relation. (B) At a critical indentation $\xi_c\sim h/R$ the capsid
    partially inverts its shape. }
\end{figure}

The elastic behavior of spherical capsids under axial point force can
be understood by simple scaling arguments \cite{landau} of an elastic
sphere with energy $E=E_e+E_b$. For small indentations $X$ the top of
the sphere is flattened in a circular region of diameter $d$, see Fig.
\ref{fig:4}A.  In the flattened region the radius of curvature $\sim
d^2/X$ and meridians are compressed by an amount $\sim\xi$.  In
equilibrium one then finds $d\sim R/\gamma^{1/4}$ yielding a total
energy $E/\kappa_b\sim\gamma^{1/2}\xi^2$ of the deformed sphere.
Thus, the force-distance relation is linear with $K_l\sim
\gamma^{1/2}$.

In fact, a more rigorous analysis of the linear regime yields
\cite{buenemann_tbp} Eq. (\ref{eq:4}) for $\gamma>10$, see Fig. 1 in
the supplementary material.  For $\gamma<10$ the scaling picture
breaks down since the dominant bending energy is no longer
concentrated in a small region around the pole.

At a critical indentation $\xi_c\sim h/R\sim \gamma^{-1/2}$ the shell
undergoes a larger shape transformation in which a circular region in
the upper part inverts its shape, see Fig.  \ref{fig:4}B. Here, the
deformation energy is concentrated in a ring of radius $r$ and width
$d$.  The radius of curvature of the ring is $\sim R d/r$ and
meridians are shortened by a factor $\sim r d/R^2$.  By using again
$d$ to equilibrate bending and elastic energy (i.e.  $d\simeq
R/\gamma^{1/4}$) one finds $E/\kappa_b\sim\xi^{3/2} \gamma^{1/4}$
leading to a nonlinear force-distance relation $F\sim \xi^{1/2}$.

For sufficiently large $\gamma$ the inversion transition of pentamers
(shown in Fig. \ref{fig:4}B) is a first order transition (in $\xi$) as
can be seen from the $F(\xi)$-curve in Fig.  \ref{fig:1}C.  There,
only the green lines correspond to conformations of minimal energy.
The blue lines represent intermediate (metastable or unstable)
conformations connecting the weakly and strongly deformed capsid
shapes. For smaller $\gamma$ the transition is continuous, see Fig.
\ref{fig:1}B.

With our numerical simulations it is also possible to extract precise
material parameters of experimentally investigated viruses.  To do so
we have simulated the conditions corresponding to the SFM experiments
on $\phi29$ \cite{wuite04}.  By using the FvK number $\gamma$ as fit
parameter for the measured bimodality ratio
$K_l^\mathrm{pent}/K_l^\mathrm{hex}\simeq 1.88$ one finds
$\gamma\simeq 1778$, in good agreement with the value estimated above from
the geometrical parameters of Ref.  \cite{tao98}.

Then, $\kappa_b/R^2 \simeq 1.14$mN/m can be extracted directly from
the dimensionless spring constant $K_l \equiv k_l R^2/\kappa_b$ shown
in Fig.  \ref{fig:3} by using, e.g., the softest experimentally
measured value of $k_l=0.16$N/m (with corresponding
$k_l^\mathrm{pent}=0.296$N/m) \cite{wuite04}.  The force scale is then
$\kappa_b/R \simeq 24$pN.  For empty CCMV the bimodality cannot be
resolved experimentally ($k_l=0.15\pm0.01$N/m \cite{wuite06}), while
numerically we find $k_l^\mathrm{hex}=0.17$N/m and
$k_l^\mathrm{pent}=0.12$N/m.  In this case, we extract $\kappa_b/R^2
\simeq 2.4\pm0.2$mN/m and $\kappa_b/R\simeq 29\pm2$pN from the
numerical data.

\begin{figure}
  \begin{center}
    {\epsfig{figure=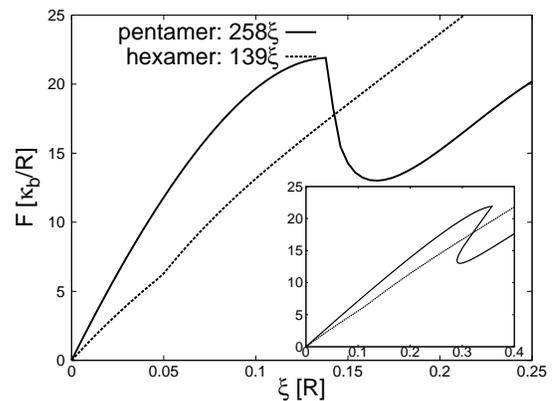,width=7.1cm}} 
  \end{center}
  \caption{\label{fig:g3.25_all} Numerical indentation experiment on
    $\phi$29 (with $\gamma=1778$). At an indentation of $0.13R$ the
    mechanical response of the pentamers suddenly decreases due to
    shape inversion, see Fig. \ref{fig:4}B.  Linear spring constants
    are determined by fitting the region of small $\xi$, implying
    $F=258\xi$ for pentamers and $F=139\xi$ for hexamers.  The inset
    displays the same curve as it would be measured in an experiment
    with an elastic cantilever with spring constant
    $K_c=100$.  }
\end{figure}

With these material parameters a direct comparison with SFM
experiments can be achieved. Fig. \ref{fig:g3.25_all} shows as an
example the force-distance curves for hexamers and pentamers of $\phi
29$. Within our numerical analysis it is even possible to take into
account the elastic spring constant of the cantilever $K_c$ (measured
in units of $\kappa_b/R^2$). Then, a measured force $F$ corresponds to
a displacement $X'/R=F/K_c + \xi$ of the shell.  Thus, as Fig.
\ref{fig:g3.25_all} shows for sufficiently stiff cantilevers the shape
inversion of the viral shell leads to a discontinuous
force-displacement curve.  The numerical results (shown in the inset)
are in excellent agreement with the experimentally measured
force-distance curves \cite{wuite04}.  In particular, for both
numerical and experimental deformations the shape instability occurs
at $\xi_c\simeq 0.13$

{\bf Rupture.} We have also analyzed the {\em local} response of
elastic shells to loading by investigating the conditions under which
rupture occurs.  This process takes place in regions of high in-plane
stress where bonds break due to over-stretching or compression.  Thus,
this analysis provides information about the interaction between
capsomers.

As experimentally observed, phage $\phi$29 breaks at a critical
indentation $X_r\simeq7.7 \mathrm{nm}\simeq 0.35R$ \cite{wuite04}.  To
determine the associated change in bond length we have calculated
numerically the shape of a shell with $\gamma=1778$ at this
indentation yielding a maximal compression of $\simeq4.5\%$.  The
corresponding numerical rupture force is
$f_r\simeq43\kappa_b/R\simeq1.0$nN in agreement with experimental
findings \cite{wuite04}.  Using this criterion, the numerics for empty
CCMV predicts a rupture force $f_r\simeq22\kappa_b/R\simeq0.57$nN
while the experimentally determined value is $f_r\simeq0.63$nN
\cite{wuite06}.

\begin{figure}
  \begin{center}
    {\epsfig{figure=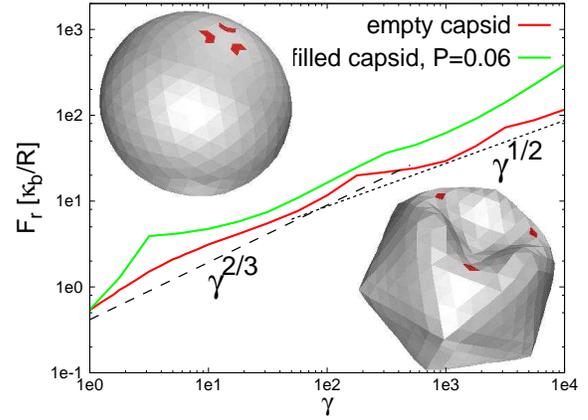,width=7.5cm}} 
  \end{center}
  \caption{\label{fig:lengthV_all} (Dimensionless) rupture force of an
    icosahedral capsid with (green curve) and without (red curve)
    volume constraint.  Rupture is assumed to occur at a critical bond
    length change of $4.5\%$.  The rupture force increases with
    $\gamma$ since it becomes harder to stretch/compress the material.
    The rupture force shows a $\gamma^\alpha$ dependence. We find
    $\alpha=0.65\pm0.03\simeq2/3$ for low $\gamma$ and
    $\alpha=0.53\pm0.04\simeq1/2$ for large $\gamma$.  The crossover
    takes places at the inversion transition shown in Fig.
    \ref{fig:4}.  The insets show ruptured structures for $\gamma=10$
    and $\gamma=3162$.  The bonds where rupture occurs first are shown
    as red diamonds.}
\end{figure}

In the following, we analyze rupture of viral shells as a function of
$\gamma$.  Fig. \ref{fig:lengthV_all} shows the $\gamma$-dependence of
$f_r$. For $\gamma < 100$ rupture occurs in the linear regime, see
Fig. \ref{fig:4}A.  Here, rupture is caused by compression of
meridians, see left inset in Fig. \ref{fig:lengthV_all}.  In the
linear regime, we observe that the rupture force scales approximately
$\sim \gamma^{2/3}$.  However, most viruses have higher FvK numbers
and will therefore rupture in the nonlinear regime after shape
inversion, see Fig.  \ref{fig:4}B.  Here, rupture is caused by
circumferential compression in the highly bent rim.  In this regime,
the rupture force scales (roughly) $\sim \gamma^{1/2}$.

We have also investigated the influence of the capsid geometry on
rupture by comparing an icosahedral and a sphero-cylindrical capsid.
This is motivated by the fact that even-numbered $T$-phages and
$\phi29$ have additional capsomers in the equatorial region elongating
the icosahedral shape. For example, the head of $T4$ is an icosahedron
with triangulation number 13 and one additional ring of hexamers
\cite{fokine04}.  The geometry of the sphero-cylinder used in our
simulations is that of $\phi$29, i.e.  an icosadeltahedron with
triangulation number $3$, elongated by two additional equatorial rings
of radius $R$ \cite{tao98}.  The force was applied perpendicularly to
the axis of symmetry.  In order to avoid tilting of the structure, the
lowest 10\% of vertices of the shell were kept fixed in the
simulations.

\begin{figure}
  \begin{center}
    {\epsfig{figure=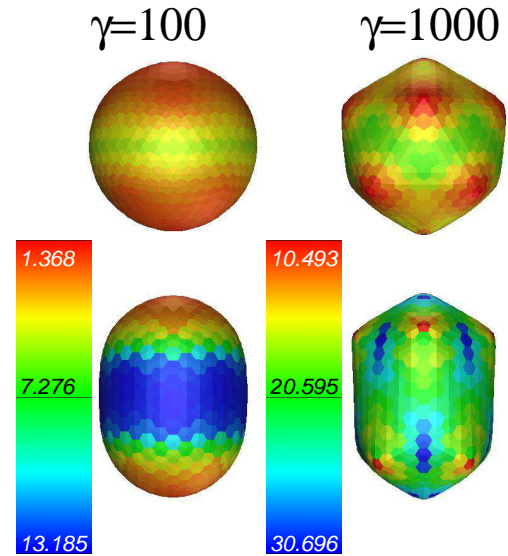,width=7.5cm}} 
  \end{center}
  \caption{\label{fig:rupfrc} Rupture force in units of $\kappa_b/R$
    plotted across the surface of spherical and sphero-cylindrical
    capsids with $\gamma=100$ and $\gamma=1000$. Rupture is assumed to
    occur at a critical bond length change of $4.5\%$.  }
\end{figure}

Fig. \ref{fig:rupfrc} shows the rupture force mapped across the
surface for $\gamma=100$ and $\gamma=1000$. In the corresponding
numerical simulations every vertex was indented in steps of $R/250$
followed by the calculation of the new equilibrium conformation.  When
stretching of some bond exceeded 4.5\% the applied force was
determined.  For $\gamma<\gamma_b$ below the buckling threshold the
shells have a uniform elastic behavior across the surface.  Since the
force exerted on the caps has a tangential component, part of the
displacement directly leads to bond stretching.  Therefore, the caps
are the regions of highest instability.  Above the buckling threshold,
the pentamers are rigid cones but the space between them becomes more
flexible.  When pushing a pentameric region the displacement is mainly
transformed into compression of the ridges.  Pentamers are thus the
most unstable region. On average a larger (scale-free) force is
required to break the capsid for larger FvK number $\gamma$.

\begin{figure}
  \begin{center}
    {\epsfig{figure=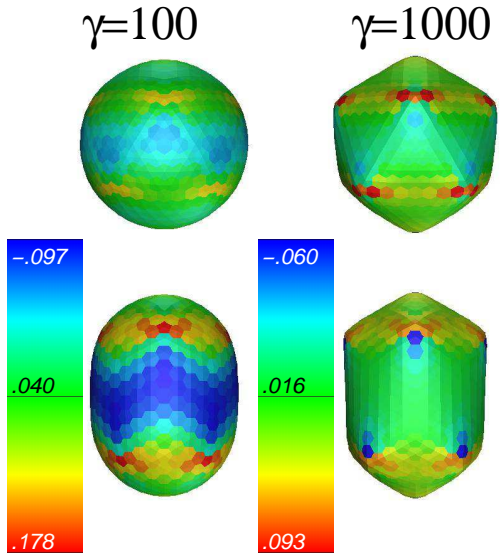,width=7.5cm}} 
  \end{center}
  \caption{\label{fig:tot} Averaged distribution of elastic
    deformation energy for an ensemble measurement where every vertex
    was indented by an amount $\xi=0.1$. In the simulations, loading
    was perpendicular to the axis of symmetry.  The ensemble average
    shows that the elastic energy is concentrated in a small region
    just below the caps.  The capsid will most likely break in this
    region.  }
\end{figure}

The spatial distribution of deformation energy can serve as a measure
for the tendency to rupture.  To mimic an ensemble of indentation
experiments, we carried out simulations in which each vertex of the
shell was indented by a constant $\xi$ ($\xi=0.1$).  For all vertices
the deformation energy was recorded for all deformations.  In order to
find the spatial distribution of deformation energy in the ensemble,
the elastic energy per vertex was averaged over the ensemble
measurement.  Fig.  \ref{fig:tot} shows a map of the average spatial
distribution of elastic energy, normalized by the total elastic energy
put into the system during the measurement on the ensemble.

In our simulations we assume a uniform distribution of binding
energies between capsomers. In reality, binding energies will show a
spatial variation and an ensemble of indentation experiments on
capsids will yield different rupture probabilities than the numerical
data presented in Fig.  \ref{fig:tot}.  However, with a direct
comparison of the two approaches conclusions about the distribution of
binding potentials can be drawn.

{\bf Filled Capsids.} Phages infect their host cells by penetrating
the cell membrane and rapidly injecting the DNA into the cell plasma.
Therefore, their DNA is stored under high pressure.  The pressure $p$
of DNA inside $\phi29$ is of the order of $p\simeq$6MPa
\cite{smith01,purohit03} or in units of elastic parameters
$p\simeq0.06\kappa_e/R$.

Other viruses, like CCMV, self-assemble inside the host cell and
enclose the DNA.  Correspondingly, their internal pressure is much
lower than that of phages. Typically, $p=$1atm$=0.01\kappa_e/R$ is a
good estimate for the internal pressure of a filled CCMV capsid.

To investigate the influence of internal pressure on the elastic
properties we have numerically simulated filled capsids with the
methods described above.  However, here the additional constraint has
to be taken into account that due to the presence of DNA inside the
capsids the enclosed volume is fixed.  In order to mimic a packed
$\phi29$ phage, first the equilibrium conformation under pressure was
determined by minimizing the total energy with an additional pressure
contribution $E_p=pV$ with $p=0.06\kappa_e/R$.

Experiments on full CCMV were performed in Ref.  \cite{wuite06}. A
direct comparison of our numerical simulations with these experiments
yields (for $p=0.01\kappa_e/R$ and volume $V=$const.) the spring
constants $k_l^\mathrm{hex}\simeq0.19$N/m,
$k_l^\mathrm{pent}\simeq0.12$N/m, and a rupture force
$f_r^\mathrm{full}\simeq0.83$nN.  These values are again in good
agreement with the experimental findings ($k_l=0.20\pm0.02$N/m and
$f_r^\mathrm{full}=0.81\pm0.04$nN).  There are no SFM-experiments on
filled $\phi$29 phages to which we could compare our numerical
simulations.

Generally, the volume constraint leads to increased linear spring
constants, see Fig. \ref{fig:3}.  This can be understood in the
scaling picture: In order to preserve constant cross-sectional area,
the local compression at the point of loading must be compensated by
an expansion of the equatorial area.  Therefore the influence of the
volume constraint becomes more apparent for larger $\gamma$, i.e. for
shells dominated by elastic energy.

At high pressures the circumferential stress at the fivefold
disclinations is balanced by the volume contribution $pV$.  Thus,
pentamers do not form rigid buckles and hexamers remain flat.  For
example, the DNA pressure of $\phi$29-phages reduces its aspherity $a$
from $a\simeq 1.4\cdot 10^{-3}$ for an empty capsid to $a\simeq 0.6
\cdot 10^{-3}$.  This similarity between hexamers and pentamers is
reflected by the fact, that the elastic response of hexamers and
pentamers becomes similar for high internal pressure, see Fig.
\ref{fig:3}.

The rupturing behavior of filled capsids is shown in Fig.
\ref{fig:lengthV_all}. The rupture force is larger for filled capsids
than for empty ones. Due to the internal pressure the non-indented
shell is already stretched. This compensates the force-induced
compression in meridional direction at the poles thus reducing their
tendency to rupture. Rupture of filled capsids is caused by
circumferential expansion at the equator.

{\bf Osmotic Shock.} One way to extract DNA from viral capsids is to
put them under osmotic shock. Under these conditions some viral
capsids (e.g. $T$-even phages) rupture \cite{anderson50} while others
(e.g. $T$-odd phages) stay intact.  Thus, these experiments also
elucidate details about capsomer-capsomer interactions and we have
analyzed whether rupture induced by internal pressure and by external
force are related.

\begin{figure}
  \begin{center}
    {\epsfig{figure=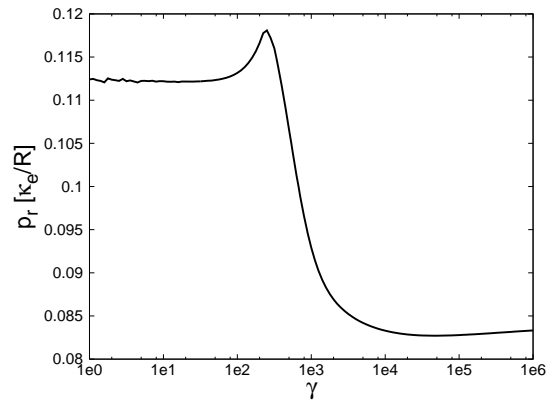,width=7.1cm}}
  \end{center}
  \caption{\label{fig:prupture} Rupture pressure (i.e. pressure at
    which the relative bond stretching reaches 4.5\%) for an
    icosahedral shell as function of $\gamma$.
  }
\end{figure}

To do so we have numerically determined the shape of the capsid under
internal pressure $p$ and measured the bond length between neighboring
vertices.  It was assumed that at rupture pressure the critical bond
stretching is 4.5\% as for rupturing due to bond compression.  Fig.
\ref{fig:prupture} shows the corresponding rupture pressure $p_r$ (in
units of $\kappa_e/R$) as function of $\gamma$ for an icosahedral
shell.

As can be seen from Fig.  \ref{fig:prupture}, capsids with a buckled
ground state ($\gamma>\gamma_b$) rupture at lower $p_r$ than unbuckled
ones. One should note, that for rupture induced by an external force,
rupture forces are generally higher for $\gamma>\gamma_b$ than for
$\gamma<\gamma_b$, see Fig. \ref{fig:lengthV_all}.  However, as can be
seen from the inset in Fig.  \ref{fig:lengthV_all}, in the buckled
configuration the force-induced deformation propagates over a larger
area than in the unbuckled state. Therefore, it is possible that the
external rupture pressure for force-induced rupture is also lower for
buckled capsids.  But this questions needs to be addressed in a more
detailed analysis.

For small $\gamma$ the rupture pressure reaches a constant value, see
Fig.  \ref{fig:prupture}. In this limit deviations from spherical
shape are small and the internal pressure simply leads to a
form-invariant up-scaling of the viral shell. Thus, changes in bending
energy can be neglected. By equilibrating the expanding pressure force
and the restoring force arising from bond stretching for a
triangulated sphere the relative change of bond length is found to be
$pR/(3\kappa_e)$. The numerically found value is somewhat lower caused
by the small deviations from spherical shape. The same calculation
shows that a triangulated sphere with a harmonic stretching energy is
only stable for pressures $p <0.75\kappa_e/R$, independently of
$\gamma$, see supporting information.

{\bf Summary and Conclusions.} Recent SFM-measurements have revealed
the elastic properties and mechanical limits of viral capsids.  In
this paper, we have shown that numerical simulations are a powerful
tool in complementing these experimental efforts. We have examined the
reversible and irreversible mechanical behavior on local and global
scales.  In particular, we have shown that the elastic parameters
characterizing the mechanical properties of phages and viruses can be
determined very precisely by a direct comparison of numerical and
experimental data. With the numerical methods presented here, one is
able to resolve the mechanical response of viral shells to an external
force in high detail. In particular, since we are able to distinguish
the response of hexamers and pentamers we can identify the origin of
the experimentally observed bimodality of the elastic spring
constants. We also make predictions for the elastic response and
rupturing behavior of both empty and filled capsids as function of the
FvK number $\gamma$.  Furthermore, in our simulations the mechanical
limits of viral shells can be probed in an ensemble where every
capsomer is indented. The comparison of the corresponding rupture map
of the shells with experimental data on SFM-induced rupturing offers
new methods in experimentally probing the local protein-protein
interactions. We can also use the rupture criterion to predict the
maximal sustainable internal pressure of capsids.

So far, we have focused on shells with the shape of an
icosadeltahedron. However, many viruses such as phage HK97
\cite{helgstrand03} or the animal virus polyoma \cite{stehle96} are
chiral. The influence of chirality on the elastic properties and
mechanical stability will be addressed in a forthcoming publication
\cite{buenemann_tbp}.

{\bf Acknowledgment.}  PL gratefully acknowledges support through the
Fonds der Chemischen Industrie and the Center for Theoretical
Biological Physics (National Science Foundation Grants Nos. PHY0216576
and PHY0225630).

\bibliographystyle{pnas}

\bibliography{./paper}


\end{article}

\end{document}